\documentclass[reprint,twocolumn,amsmath,amssymb,aps,superscriptaddress,prl]{revtex4-1}
\usepackage{graphicx}
\usepackage{bm}
\usepackage{epsfig}
\usepackage{amssymb}
\usepackage{amsfonts}
\usepackage{amsmath}
\usepackage{braket}
\usepackage{color}
\usepackage{epstopdf}
\usepackage{hyperref}
\usepackage{float}
\restylefloat{table}
\usepackage{bibentry}
\usepackage{multirow}
\usepackage[caption=false]{subfig}
\usepackage{braket}
\newcommand{\ba}{\begin{eqnarray}}
\newcommand{\ea}{\end{eqnarray}}
\newcommand{\bd}{\begin{displaymath}}
\renewcommand{\v}[1]{{\bf #1}}
\newcommand{\nn}{\nonumber \\}
\DeclareGraphicsExtensions{.pdf,.png,.jpg}

\begin{document}
\title{Construction of Variational Matrix Product States for the Heisenberg Spin-1 Chain}

\author{Jintae Kim}
\affiliation{Department of Physics, Sungkyunkwan University, Suwon 16419, Korea}
\author{Minsoo Kim}
\affiliation{Department of Physics, Sungkyunkwan University, Suwon 16419, Korea}
\author{Naoki Kawashima}
\affiliation{Institute for Solid State Physics, University of Tokyo, Kashiwa, Chiba 277-8581, Japan}
\author{Jung Hoon Han}
\email[Electronic address:$~~$]{hanjemme@gmail.com}
\affiliation{Department of Physics, Sungkyunkwan University, Suwon 16419, Korea}
\author{Hyun-Yong Lee}
\email[Electronic address:$~~$]{hyunyong@korea.ac.kr}
\affiliation{Department of Applied Physics, Graduate School, Korea University, Sejong 30019, Korea}
\affiliation{Division of Display and Semiconductor Physics, Korea University, Sejong 30019, Korea}
\date{\today}
\begin{abstract} We propose a simple variational wave function that captures the correct ground state energy of the spin-1 Heisenberg chain model to within 0.04\%. The wave function is written in the matrix product state (MPS) form with the bond dimension $D=8$, and characterized by three fugacity parameters. The proposed MPS  generalizes the Affleck-Kennedy-Lieb-Tasaki (AKLT) state by dressing it with dimers, trimers, and general $q$-dimers. The fugacity parameters control the number and the average size of the $q$-mers. Furthermore, the $D=8$ variational MPS state captures the ground states of the entire family of bilinear-biquadratic Hamiltonian belonging to the Haldane phase to high accuracy. The 2-4-2 degeneracy structure in the entanglement spectrum of our MPS state is found to match well with the results of density matrix renormalization group (DMRG) calculation, which is computationally much heavier. Spin-spin correlation functions also find excellent fit with those obtained by DMRG.  
\end{abstract}
\maketitle
%

{\it Introduction}: Examples of exact many-body ground states tied to relatively simple Hamiltonians are extremely rare. In the case of the Affleck-Kennedy-Lieb-Tasaki (AKLT) ground state, whose Hamiltonian is~\cite{AKLT87,AKLT88}
\ba H_{\rm A} =\sum_i \Bigl(\frac{1}{3}({\v S}_i\cdot {\v S}_{i+1})^2+ {\v S}_i\cdot {\v S}_{i+1}+\frac{2}{3}\Bigl) , \ea
the simplicity of the wave function is revealed through its matrix product state (MPS) form with the bond dimension $D=2$ - the smallest dimension allowed in any MPS representation\cite{Klumper92}. The ground state of the pure spin-1 Heisenberg model belongs to the same Haldane\,\cite{Haldane83} or the symmetry protected topological\,(SPT)\cite{Pollmann10,Xie11} phase as the AKLT state, {\it i.e.}, the two states are in some sense smoothly connected to each other. One aspect of the continuity is the double degeneracy of the entanglement spectrum (ES) that characterizes the whole Haldane phase~\cite{Pollmann10}. Beyond that, there has been little effort at establishing the continuity of the ground states within the spin-1 SPT phase at the level of wave functions themselves. We address this issue with the construction of a $D=8$ variational MPS wave function with excellent ground-state properties  for the whole family of bilinear-biquadratic (BLBQ) Hamiltonians belonging to the Haldane phase. 
\\

\begin{figure}[h]
\includegraphics[width=0.46\textwidth]{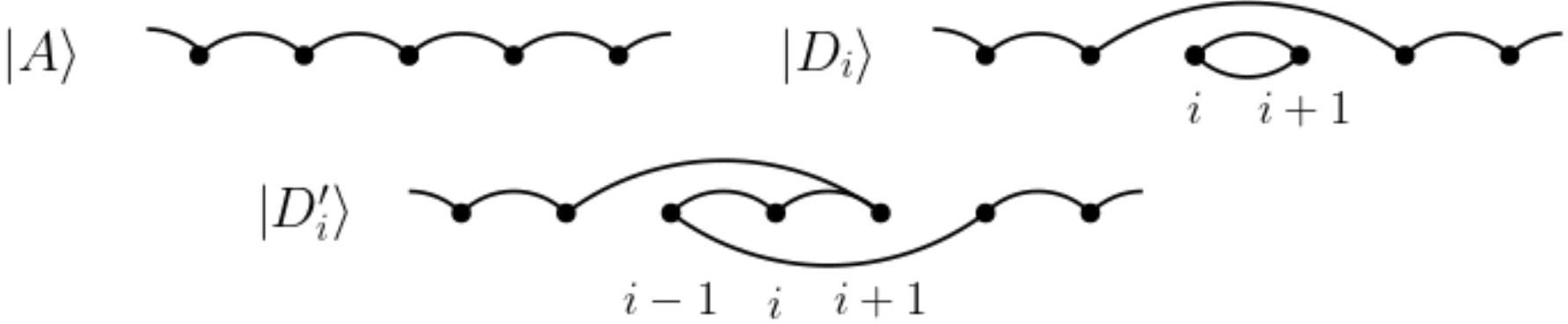} 
\caption{Schematic figures for the AKLT state $|A\rangle$, the single-dimer state $\ket{D_{i}}$,  and the extended dimer state $|D'_i \rangle$.} \label{fig:1}
\end{figure}

{\it Construction of the $D=8$ MPS wave function}: A useful way to think about the Heisenberg Hamiltonian $H_{\rm H}$ is as an AKLT model with perturbation $H_{\rm H} = H_{\rm A} + \lambda \sum_i \v S_i \cdot \v S_{i+1}$, in the limit $\lambda \rightarrow \infty$. In this view, modification to the AKLT state occurs by the action of the quadratic exchange $\v S_i \cdot \v S_{i+1}$ on the AKLT state $| A \rangle$. We find 
\ba \label{eq:first-order} 
\v S_i \cdot \v S_{i+1} | A  \rangle = -  |A \rangle + (1/2) |D_{i} \rangle .  \label{eq:dimer} 
\ea
The new state $|D_{i}\rangle$, shown in Fig. \ref{fig:1}, has a pair of adjacent sites $(i,i+1)$ locked into the total spin-0 ``dimer", while the rest of the sites remains in the AKLT state. In the Schwinger boson (SB) notation for the singlet creation operator $s^\dag_{ij} = a^\dag_i b^\dag_j - b^\dag_i a^\dag_j$, 
\ba \ket{D_i } \!=\! \Bigl[ \bigl( \prod_{j\neq i-1, i, i+1} s_{j,j+1}^{\dagger} \bigr) s_{i-1,i+2}^{\dagger}  \Bigr]   (s_{i,i+1}^{\dagger})^2 \ket{v} \nonumber
\ea 
with $|v\rangle$ being the SB vacuum. The terms inside the bracket $[ \cdots ]$ give the AKLT state over the chain with two sites missing. The appearance of an isolated dimer according to Eq. (\ref{eq:dimer}) was noted by Arovas some time ago~\cite{arovas89}. 

According to the first-order consideration above, the ground state of the Heisenberg model differs from the AKLT state by the appearance of a single dimer. At the next order in perturbation we find 
\ba\label{eq:second-order}
{\v S}_i\cdot{\v S}_{i+1}\ket{D_{i}}&=&-2\ket{D_{i}}\nn
{\v S}_{i-1}\cdot{\v S}_{i}\ket{D_{i}}&=&\ket{A}+\ket{D_{i}}+\ket{D'_i}\nn
{\v S}_{i+1}\cdot{\v S}_{i+2}\ket{D_{i}}&=&\ket{A}+\ket{D_{i}}+\ket{D'_{i+1}}\nn
{\v S}_j\cdot{\v S}_{j+1}\ket{D_{i}}&=&-\ket{D_{i}}+\frac{1}{2}\ket{D_i D_j},
\ea
where $j \neq i-1, i, i+1$ in the last line. The new state $\ket{D'_i}$, shown in Fig. \ref{fig:1}, can be decomposed as the superposition of the AKLT state, one-dimer states $| D_j\rangle$, and a new, length-2 dimer state defined over the second-nearest neighbors $(i,i+2)$. The appearance of the extended, length-2 dimer is the new feature of the second-order perturbation along with the double-dimer configuration $|D_i D_j \rangle$ at the non-overlapping bonds $(i,i+1)$ and $(j,j+1)$. More details on the algebra of dimers can be found in the Supplementary Material (SM).  The perturbative considerations are useful guides in anticipating what new configurations characterize the ground state of the spin-1 Heisenberg Hamiltonian. 

\begin{figure}[h]
	\includegraphics[width=0.49\textwidth]{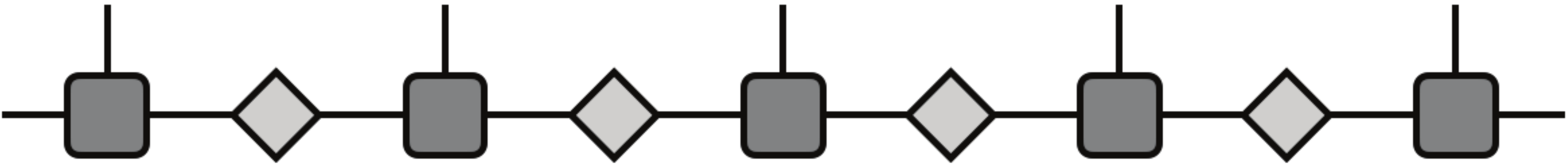} 
	\caption{General graphical representation of the MPS consisting of alternating site tensor (square with a vertical arm) and bond tensor (diamond, no dangling arm). 
	} 
	\label{fig:2}
\end{figure}

The MPS wave function is given by the product of alternating site and bond tensors as depicted in Fig. \ref{fig:2}. Certain constraints, such as the SO(3) symmetry of the state, must be met in the choice of tensors (more on the symmetry discussion in SM). The $D=8$ site tensor we propose consists of the sum
\ba
T = T_1 + T_2  + 4 a b^2 \, T_2^t - 2b \,T_3  - 3c \, T_4 \label{eq:site-tensor}
\ea
involving three free parameters $a,b,c$. The components of the tensors are given out explicitly, 
\ba
	 \left[T_1\right]_{ii',jj'}^s &= & \left[ {\rm CG}_{\frac{1}{2}\frac{1}{2}}^1\right]_{ij}^s \delta_{i'3} \delta_{j'3} \nn
	 \left[T_2\right]_{ii',jj'}^s &= &\left[ {\rm CG}_{\frac{1}{2}\frac{1}{2}}^0\right]_{ij} \delta_{i'3} \delta_{j' s} \nn
	 \left[T_3\right]_{ii',jj'}^s &=& \left[ {\rm CG}_{\frac{1}{2}\frac{1}{2}}^0\right]_{ij}\left[ {\rm CG}_{11}^1\right]_{i'j'}^s \nn 
	 \left[T_4\right]_{ii',jj'}^s &=& \left[ {\rm CG}_{\frac{1}{2}\frac{1}{2}}^1\right]^s_{ij}\left[ {\rm CG}_{11}^0 \right]_{i'j'} \nn 
	 \left[B\right]_{ii',jj'} &=& \left[ {\rm CG}_{\frac{1}{2}\frac{1}{2}}^0\right]_{ij} \bigl( \left[ {\rm CG}_{11}^0\right]_{i'j'} + \delta_{i'3} \delta_{j' 3} \bigr) .
	 \label{eq:tensor-components}
\ea
The bond tensor $B$ is shown in the last line. The $T^t_2$ in Eq. (\ref{eq:site-tensor}) is obtained by taking the transpose of the virtual indices, $[T_2^t ]^s_{ii',jj' }   = [T_2 ]^s_{jj',ii' }$. Clebsch-Gordon (CG) coefficients ${\rm CG}_{s_1 s_2}^s$ for combining two spins $s_1$ and $s_2$ into the spin $s$ are employed above. In the case of ${\rm CG}_{11}^0$ and ${\rm CG}_{11}^1$, the virtual indices $i', j'$ run only over the three possible spin states $0, 1, 2$. Nevertheless we introduce a fourth component $i' , j' = 3$ and make them four-dimensional. This mathematical contraption plays a crucial role in our construction. Meanwhile the un-primed indices are two-dimensional, $i, j = 0,1$, for a total of $2\times 4 = 8$-dimensional virtual indices, or $D=8$. The wave function for a given spin basis $|s_1 s_2 \cdots s_N \rangle$ is obtained by taking the tensor product $T^{s_1} B T^{s_2} B \cdots T^{s_N} B$ and contracting the two end indices either with a trace or some boundary vectors.

The product of a site tensor $T^s$ with the adjoining bond tensor $B$ is symbolically written $T B = \overline{T} $. In components,
\ba 
	 [ \overline{T}_1 ]_{ii',jj'}^s &= & \left[ {\rm CG}_{\frac{1}{2}\frac{1}{2}}^1\right]_{ik}^s  \left[ {\rm CG}_{\frac{1}{2}\frac{1}{2}}^0\right]_{kj}  \delta_{i'3} \delta_{j'3} \nn
	 \left[ \overline{T}_2 \right]_{ii',jj'}^s &= &(-\delta_{ij} /2) \delta_{i'3} \left[ {\rm CG}_{11}^0\right]_{sj'} \nn
         \left[ \overline{T}_2^t \right]_{ii',jj'}^s &= &(\delta_{ij} /2) \delta_{i's} \delta_{j' 3} \nn
          \left[ \overline{T}_3 \right]_{ii',jj'}^s &=& (-\delta_{ij}/2) \left[ {\rm CG}_{11}^1\right]_{i'k'}^s \left[ {\rm CG}_{11}^0\right]_{k' j' }  \nn
	  \left[ \overline{T}_4 \right]_{ii',jj'}^s &=& (\delta_{i'j'}/3)  \left[ {\rm CG}_{\frac{1}{2}\frac{1}{2}}^1\right]_{ik}^s  \left[ {\rm CG}_{\frac{1}{2}\frac{1}{2}}^0 \right]_{kj} .  
	 \label{eq:tensor-components2}
\ea
The two relations $ [{\rm CG}_{\frac{1}{2}\frac{1}{2}}^0]_{ik} [ {\rm CG}_{\frac{1}{2}\frac{1}{2}}^0]_{kj} = -\delta_{ij}/2$ and 
$[ {\rm CG}_{11}^0 ]_{i' k'} [ {\rm CG}_{11}^0]_{k' j'} = \delta_{i'j'}/3$ were used.  Summations over repeated indices are implicit. Note that 
$\left[ {\rm CG}_{\frac{1}{2}\frac{1}{2}}^1\right]_{ik}^s  \left[ {\rm CG}_{\frac{1}{2}\frac{1}{2}}^0\right]_{kj}   \equiv A^s_{ij}$ is precisely the MPS tensor that defines the AKLT state. In the simplest case $a=b=c=0$, the product of $T_1$ tensors reproduces the AKLT state. 

\begin{figure}[!h]
\includegraphics[width=0.49\textwidth]{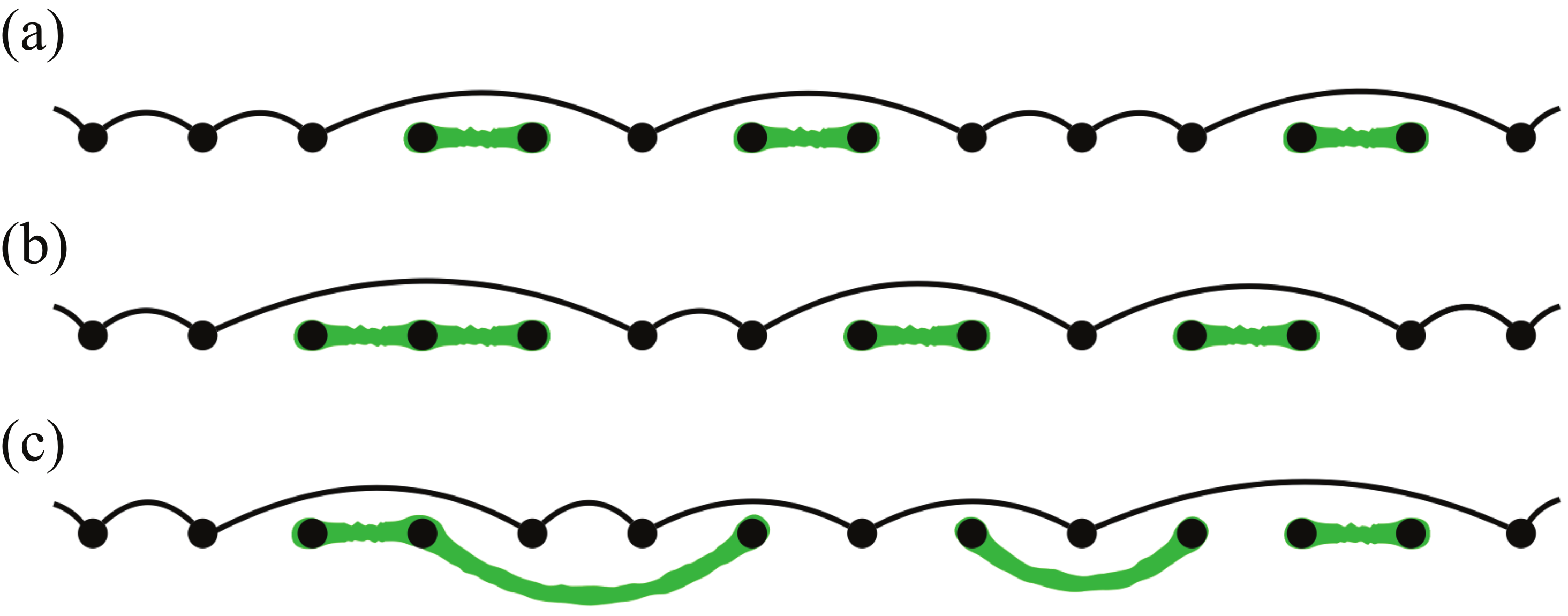} 
\caption{Exemplary configurations containing (a) multiple compact dimers,  (b) two dimers and one trimer (all compact) and (c) long-ranged $q$-mers. The black solid line stands for the singlet made out of two $S=1/2$'s, while the thick green ones are the dimers and trimers. } 
\label{fig:3}
\end{figure}

Next, we keep $T_1, T_2$ and its transpose and examine the resulting MPS state. From the tensor structure shown in Eq. (\ref{eq:tensor-components2}), one finds that $\overline{T}_2$ can only be followed by $\overline{T}_2^t$ and not by $\overline{T}_1$. This constraint effectively binds the $\overline{T}_2$ and its transpose into a pair,
\ba  [ \overline{T}_2 ]_{ii' , kk' }^s [ \overline{T}_2^t  ]^{s'}_{kk', jj' } = (- \delta_{ij} /4 ) \delta_{i'3}\delta_{j' 3} [{\rm CG}_{11}^0 ]_{s s'} . \nonumber \ea 
The expression $[{\rm CG}_{11}^0 ]_{s s'}$ is nothing but the wave function of a dimer singlet. The factor $(-1/4)$ in the above combines with the prefactor $4 ab^2$ in Eq. (\ref{eq:site-tensor}) to give the factor $-ab^2$ to the one-dimer configuration $|D_i \rangle$ depicted in Fig. \ref{fig:1}. $\overline{T}_2^t$ can be followed either by $\overline{T}_2$, creating a second dimer in succession to the first, or by $\overline{T}_1$, terminating the dimer and restoring the AKLT chain. The expansion of the tensor product (still omitting $T_3$ and $T_4$) gives out the series
\begin{align}
	|\psi_{\rm DG} \rangle = \sum_{n = 0}^\infty \sum_{\Gamma_{\rm DG}^{(n)}}  (-a)^{n} b^{2n} | \Gamma_{\rm DG}^{(n)}  \rangle, \label{eq:dimer-expansion} 
\end{align}
where the sum $n$ spans the number of dimers, and $\Gamma_{\rm DG}^{(n)}$ refers to all possible arrangements of the $n$ dimers ($n=0$ gives the AKLT state). The two exponents in $(-a)^n b^{2n}$ count the number of dimers ($n$) and the total length of the dimers ($2\times n = 2n)$, respectively. For the same fugacities, {\it i.e.} the same $n$, one has all dimer configurations contributing with equal weight to the above sum - a situation we refer to as the dimer gas\,(DG). An example of the multi-dimer configuration is shown in Fig. \ref{fig:3}(a). The one-dimer configurations in the above sum contributes with a minus sign $-a b^2$, in accordance with the prediction of the first-order perturbation. Numerical minimization of the MPS energy indeed proves that $a>0$ for the variational ground state. Note that all the dimers appearing in the multi-dimer configuration in Eq. (\ref{eq:dimer-expansion}) are defined over the nearest neighbors, {\it i.e.} the dimers are ``compact". 

Next we restore $T_3$ but not yet $T_4$. In addition to the dimer-giving product $\overline{T}_2 \overline{T}_2^t$ already discussed, the product $\overline{T}_2 ( \overline{T}_3 )^m \overline{T}_2^t$ with any number of $m$'s is possible. An explicit calculation gives 
\ba && [\overline{T}_2 \overline{T}_3 \overline{T}_2^t ]^{s_1 s_2 s_3}_{ii', jj' } = (\delta_{ij} \delta_{i' 3} \delta_{j' 3} /8 )   \times \nn
&&  ~~~~~~~~~~~~ \left[ {\rm CG}_{11}^0 \right]_{s_1 \alpha} [ {\rm CG}_{11}^1 ]^{s_2}_{\alpha\beta} [ {\rm CG}_{11}^0]_{\beta s_3} . \label{eq:trimer} \ea
The local trimer wave function shown in the second line appears with the weight $-a b^3$, the exponent 3 representing the presence of a $q$-mer with $q=3$. The product $\overline{T}_2 ( \overline{ T}_3 )^2  \overline{T}_2^t$ generates the local tetramer wave function
\ba [ {\rm CG}_{11}^0 ]_{s_1 \alpha} [ {\rm CG}_{11}^1 ]^{s_2}_{\alpha\beta}  [ {\rm CG}_{11}^0 ]_{\beta\gamma }  [ {\rm CG}_{11}^1 ]^{s_3}_{\gamma\delta}  [ {\rm CG}_{11}^0]_{\delta s_4}  .  \label{eq:tetramer} \ea
The local $q$-mer is the trivial representation of the SU(2) spin rotation regardless of its length. See SM for details. One can now read off the general structure of the $q$-mer wave functions generated by the $(T_1, T_2 , T_3 )$ construction as
\ba |\psi_{\rm QG} \rangle = \sum_{\Gamma_{\rm QG}}  (-a)^{n} b^{l} | \Gamma_{\rm QG}^{(n,l)}  \rangle  . \label{eq:expansion} \ea
The symbol $\Gamma_{\rm QG}$ refers to any one of the possible mixed $q$-mer configurations. Configurations with the same total number of $q$-mers ($n$) and their total lengths given by $l = \sum_i q_i n_i$ ($q_i = 2,3$ for dimers and trimers, respectively) contribute to the wave function with the same weight, in this $q$-mer gas\,(QG) wave function $|\psi_{\rm QG}\rangle$. An example with one trimer and two dimers ($n=3, l=7$) is shown in Fig. \ref{fig:3}(b). Each $q$-mer in the expansion is still compact, or defined over $q$ consecutive sites.

As with $T_3$, the insertion of $T_4$ can only take place between $T_2$ and $T_2^t$. The role of $T_4$ is to take a compact $q$-mer and ``stretch it" over non-consecutive sites, without changing the $q$ value. To see this, include $T_1, T_2, T_4$ but not $T_3$ in the site tensor. Possible structures are $\overline{T}_2 ( \overline{T}_4 )^m \overline{T}_2^t$ with arbitrary $m$. For instance,
\ba && [\overline{T}_1 \overline{T}_2 \overline{T}_4 \overline{T}_2^t \overline{T}_1 ]^{s_1 s_2 s_3 s_4 s_5}_{ii', jj' } =  \nn
&&  ~~~~~ - (\delta_{i' 3} \delta_{j' 3} /12 )  \times  A^{s_1}_{ik} A^{s_3}_{k l } A^{s_5}_{ l j }  \left[ {\rm CG}_{11}^0 \right]_{s_2 s_4 } . \label{eq:dimer2} \ea
Indeed the dimer bond $\left[ {\rm CG}_{11}^0 \right]_{s_2 s_4 }$ is now over the second neighbors, while the AKLT tensors connect the non-adjacent sites 1,3,5. This is precisely the non-compact dimer configuration generated at the second-order perturbation as mentioned earlier. Expansion of the MPS state  (still omitting $T_3$) gives rise to the long-ranged dimer gas\,(LDG), 
\ba |\psi_{\rm LDG} \rangle = \sum_{\Gamma_{\rm LDG}} (-ab^2 )^{n} (-c)^m | \Gamma_{\rm LDG}^{(n,m)}  \rangle  . 
\ea
The number $m_i$ of insertions of $T_4$ in a given dimer gives $m= \sum_i m_i$. It is straightforward now to see that keeping all four tensors gives the expansion of the variational MPS state:
\ba |\psi_{\rm LQG} \rangle = \sum_{\Gamma_{\rm LQG}} (-a)^n b^l (-c)^m | \Gamma_{\rm LQG}^{(n,l,m)} \rangle   . \ea
Each $q$-mer has the length $q_i + m_i$. A trimer defined over the non-adjacent sites 1, 3, 5 will contribute $n=1, l = 3, m = 2$, for instance, to the weight. This picture of the long-ranged $q$-mer gas\,(LQG) sums up the nature of the variational MPS state we propose. 
\\

\begin{figure}[th]
\centering
\includegraphics[width=0.48\textwidth]{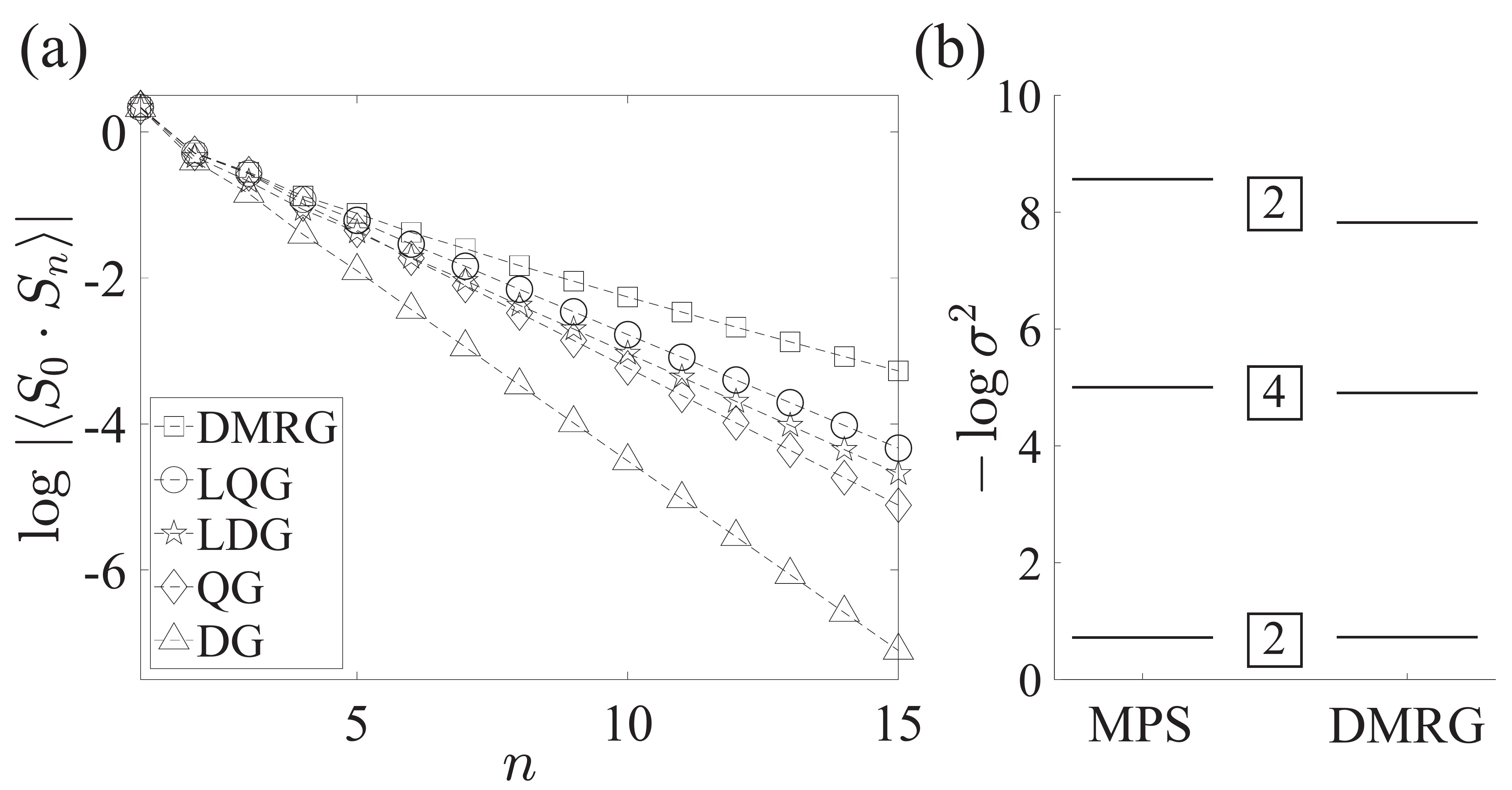} 
\caption{(a) Spin-spin correlation function $\log ( | \langle \v S_i  \cdot \v S_{i + n} \rangle | )$ (omitting the oscillatory factor $(-1)^n$) obtained from the four variational MPS states: DG, QG, LDG, and LQG. DMRG results are shown for comparison. Corresponding inverse slopes, a.k.a. correlation lengths, are 5.0940 (DMRG), 3.2143 (LQG), 3.0299 (LDG), 2.6580 (QG), 1.9249 (DG), respectively, by fitting the large-$n$ parts of the data with the linear function. (A larger correlation length of 6.03 was obtained in Ref. \cite{White93} using a different fitting procedure.) (b) Entanglement spectrum obtained from the LQG state and DMRG. The degeneracy of each level is indicated besides the levels. All variational calculations are performed to optimize the Heisenberg exchange energy, Eq. (\ref{eq:Heisenberg-E}).}
\label{fig:4}
\end{figure}

{\it Ground state of the Heisenberg model}: To test the validity of $|\psi_{\rm LQG}\rangle$ as a good variational ground state of the Heisenberg model, we calculate
\ba E(a,b,c) = \langle \psi_{\rm LQG} | \v S_i \cdot \v S_{i+1} |\psi_{\rm LQG} \rangle . \label{eq:Heisenberg-E} \ea 
Including only the ($T_1, T_2$) tensors and varying the coefficient $4ab^2$ in Eq. (\ref{eq:site-tensor}) already gives $E_{\rm DG} = -1.3920$, in good comparison to the value found by DMRG, $E_{\rm DMRG} =-1.4015$\cite{White93} and a clear improvement over the energy of the AKLT state $E=-4/3$.  Energy improves progressively with the inclusion of more tensors, $E_{\rm LDG} = -1.3991$ and $E_{\rm QG} = -1.3998$, until $E_{\rm LQG} = -1.40097$ at $(a,b,c) = (6.8990, 0.2116, 0.3564)$ becomes only 0.04\% higher than $E_{\rm DMRG}$ despite the small bond dimension $D=8$. It is remarkable that three-parameter optimization produces the comparable energy against DMRG and modern tensor network algorithms\cite{verstrate11, Zauner18} optimizing about $D^2$ parameters. A typical DMRG run employs the bond dimension $D \sim 10^{2} - 10^3$. 

The spin-spin correlation function of the LQG state, shown in Fig. \ref{fig:4}(a), is in good agreement with the DMRG results with $\langle \v S_0 \cdot \v S_n \rangle_{\rm LQG} / \langle \v S_0 \cdot \v S_n \rangle_{\rm DMRG} = 0.9996, 0.9948, 0.9852, 0.9541, 0.9085$ for $n=1,2,3,4,5$, respectively. Meanwhile, there is a significant change in the estimated correlation length $\xi$ which grows as $\xi_{\rm DG} < \xi_{\rm QG} < \xi_{\rm LDG} < \xi_{\rm LQG} < \xi_{\rm DMRG}$ as specified in the caption of Fig.\,\ref{fig:4}. The entanglement spectrum shown in Fig. \ref{fig:4}(b) displays the 2-4-2 degeneracy regardless of the $(a,b,c)$ parameters chosen, except at $a=0$ (AKLT state) where only a single pair of degenerate levels appears. The double degeneracy is the characteristic of the SPT phase protected by the $\mathbb{Z}_2 \times \mathbb{Z}_2$ spin rotation symmetry\,\cite{Pollmann10}. In fact, the virtual legs in our $D=8$ tensor accommodate the spin representation $\frac{1}{2} \otimes (0 \oplus 1)$ which is identical to $\frac{1}{2} \oplus \frac{3}{2} \oplus \frac{1}{2}$, leading to the 2-4-2 degeneracy. Furthermore, the two lowest-lying entanglement spectra from the LQG state compare favorably with those of DMRG and modern state-of-the-art algorithms\cite{verstrate11,Zauner18}:  $-\log \sigma^2 = 0.7207, 5.0060, 8.5652$ for MPS and $0.7242, 4.9045, 7.8227$ for DMRG~\cite{itensor}. 
\\

\begin{figure}[t]
\centering
\includegraphics[width=0.48\textwidth]{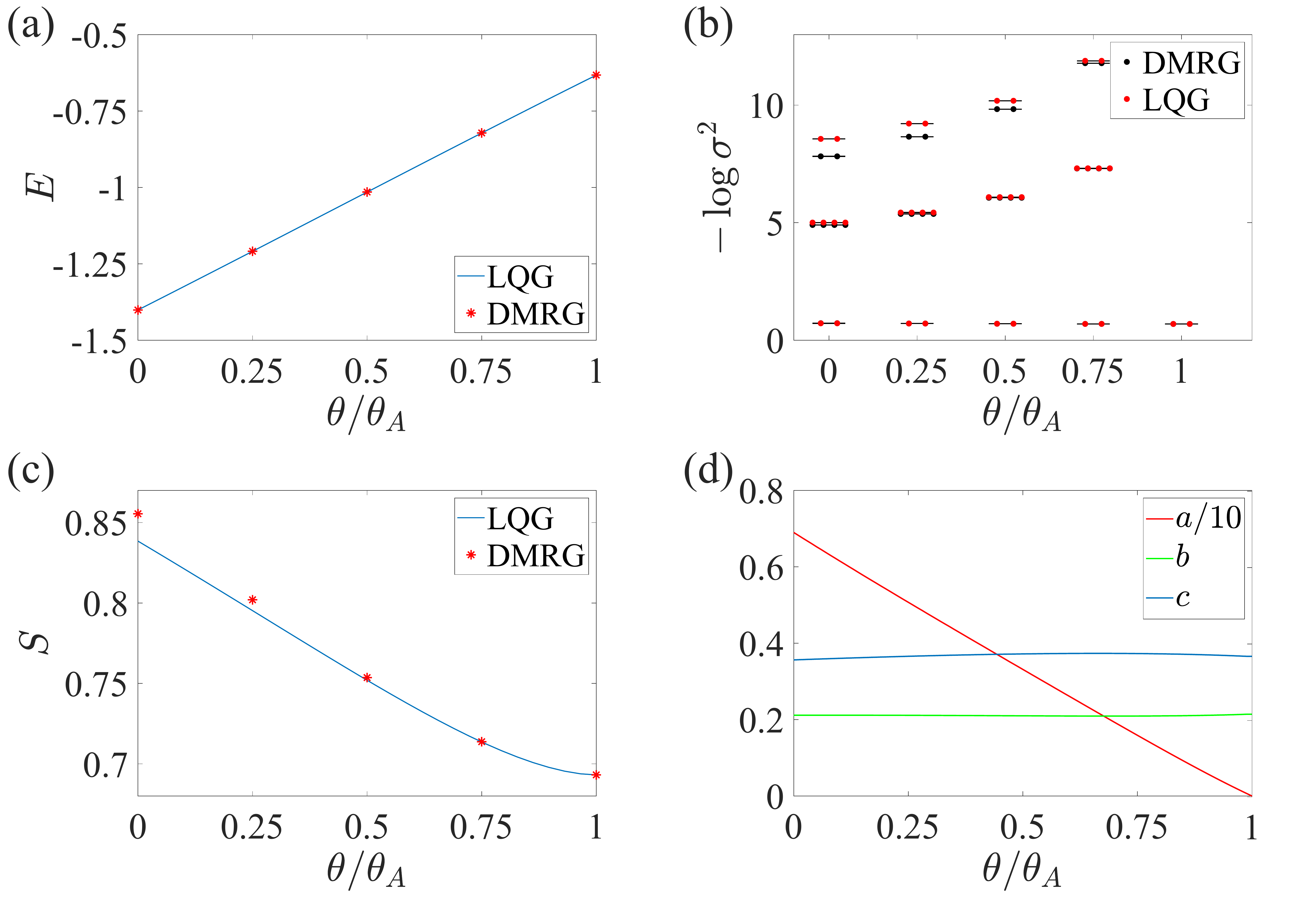}
\caption{Variational MPS optimization for the BLBQ model $H (\theta)$. (a) Optimized $(a,b,c)$ vs. $\theta$. The $a$ values have been scaled down by a factor 10 for clarity.  (b) Variational energy vs. $\theta$. Lines are from $D=8$ variational MPS after optimization, and squares are from the DMRG. Differences in energy occur in the fourth significant digits. (c) Entanglement entropy vs $\theta$. (d) Entanglement spectrum vs. $\theta$. The lowest two sets of levels agree very well between variational MPS and DMRG. }
\label{fig:5}
\end{figure}

{\it Haldane phase in the bilinear-biquadratic model}: The Heisenberg and the AKLT models are two special examples of the BLBQ spin Hamiltonian\cite{Legeza07}
\ba H(\theta) = \sum_i \left[ \cos \theta (\v S_i \cdot \v S_{i+1}) + \sin \theta (\v S_i \cdot \v S_{i+1} )^2  \right] ,\ea
with $\theta =0$ and $\theta_{\rm A}= \tan^{-1} (1/3)$ corresponding to the Heisenberg and the AKLT points, respectively. We performed optimization of the LQG for $0 \le \theta \le \theta_{\rm A}$ with various results shown in Fig. \ref{fig:4}. The spin-spin correlation data is given in the SM. The variational energy of the LQG [Fig.\,\ref{fig:5}\,(a)] shows better agreement with the DMRG as the model moves away from the Heisenberg limit towards AKLT. In addition, the entanglement spectrum and entropy are captured well all over the phase diagram as shown in Fig.\,\ref{fig:5}\,(b) and (c), respectively. The weight $a$, mainly responsible for the average number of dimers in the ground state, increases linearly from the AKLT point [Fig. \ref{fig:5}(d)]. The other parameters $b$ and $c$, having to do with the control over the average size $q$ and the spatial extent of the $q$-mer, remain nearly constant er and its extension are almost constant throughout the phase diagram. A more extensive comparison of the LQG state and the DMRG results over the whole Haldane phase $-\pi/4 < \theta < \pi /4$ is given in the SM.  
\\

{\it Discussion}: All in all, the variational MPS state with a small bond dimension $D=8$ does a good job capturing aspects of the ground states of the Haldane phase. Given the robust 2-4-2 degeneracy structure of the entanglement spectrum through the Haldane phase, $D=8$ is likely the minimum bond dimension allowed in any good MPS description of the ground state. Employing variational MPS state with even larger bond dimension will improve the accuracy of the entanglement entropy and the correlation length compared to the DMRG, at the expense of employing further variational parameters. Indeed, a theory of formal expansion of MPS tensors in terms of irreducible representation of SU(2) was developed and applied to spin-1 BLBQ model before~\cite{klumper16}. Several dozen optimization parameters were employed there, in exchange for much better numerical accuracy of the ground state energy. Our $D=8$ variational construction is developed out of the intuition obtained by perturbative consideration, and employs only three parameters while still providing reliable answer for the energy and other ground-state quantities. More importantly, it provides an intuitive picture for the character of the states in the Haldane phase as that of the AKLT parent state dressed by various compact and non-compact $q$-mers.

\acknowledgments J. H. H. was supported by Samsung Science and Technology Foundation under Project Number SSTF-BA1701-07. We acknowledge insightful comment on the manuscript from Hosho Katsura. 


\bibliography{SC}
\end{document}


\title{Supplemental Materials}

\author{Jintae Kim}
\affiliation{Department of Physics, Sungkyunkwan University, Suwon 16419, Korea}
\author{Minsoo Kim}
\affiliation{Department of Physics, Sungkyunkwan University, Suwon 16419, Korea}
\author{Naoki Kawashima}
\affiliation{Institute for Solid State Physics, University of Tokyo, Kashiwa, Chiba 277-8581, Japan}
\author{Jung Hoon Han}
\email[Electronic address:$~~$]{hanjemme@gmail.com}
\affiliation{Department of Physics, Sungkyunkwan University, Suwon 16419, Korea}
\author{Hyun-Yong Lee}
\email[Electronic address:$~~$]{hyunyong@korea.ac.kr}
\affiliation{Department of Applied Physics, Graduate School, Korea University, Sejong 30019, Korea}
\affiliation{Division of Display and Semiconductor Physics, Korea University, Sejong 30019, Korea}
\date{\today}

\maketitle

\section{Q-mer algebra}
From the calculations shown in the main text we know that 
\ba (2 \v S_i \cdot \v S_{i+1}  + 2 ) |A \rangle = |D_i \rangle . \ea
Arranged in this way, it appears that $2 \v S_i \cdot \v S_{i+1} + 2$ is playing the role of the dimer creation operator. In fact this interpretation makes sense, once we re-write the operator in the equivalent form
\ba [ (\v S_i + \v S_{i+1} )^2 - 2 ] |A \rangle = |D_i \rangle. \ea
The AKLT state $|A\rangle$ contains, by construction, only those configurations that have the total spin $S = 0$ or $S=1$ for the $(i,i+1)$ pair of sites. By acting on it with the projection operator $ (\v S_i + \v S_{i+1} )^2 - 2$, one ends up eliminating the $S=1$ components over the $(i,i+1)$ bond. The state that remains after the projection must be the total spin singlet $S=0$, which we called the dimer in the main text. 

Following a similar line of reasoning, the trimer state may be created by the operation
\ba [ (\v S_{i-1} + \v S_i + \v S_{i+1} )^2 - 2] [ (\v S_{i-1} + \v S_i + \v S_{i+1} )^2 - 6 ] | A \rangle  . \ea
Out of the three possible total spins $S=0, 1, 2$ for the $(i-1,i,i+1)$ bonds, only the $S=0$ state will survive after the projection. After expanding, we get
\ba  (\v S_{i-1} \cdot \v S_i + \v S_i \cdot \v S_{i+1} + \v S_{i+1} \cdot \v S_{i-1} )  ( \v S_{i-1} \cdot \v S_i + \v S_i \cdot \v S_{i+1} + \v S_{i+1} \cdot \v S_{i-1} + 2 ) | A \rangle = |T_i \rangle . \ea
The trimer projection operator requires the second-order action of the exchange operators on the AKLT state. This sort of reasoning is guiding us to think that $q$-mers will be generated at successively higher orders of the perturbation. 

A simple identity can be proven for the singlet creation operators $s^\dag_{ij} = a^\dag_i b^\dag_j - b^\dag_i a^\dag_j$:
\ba s^\dag_{ij} s^\dag_{kl} + s^\dag_{ik} s^\dag_{lj} + s^\dag_{il} s^\dag_{jk} = 0 . \ea
This identity can be used to prove, among others, the following relations:
\ba s^\dag_{12}s^\dag_{23} s^\dag_{34} s^\dag_{41} + s^\dag_{13}s^\dag_{34} s^\dag_{42} s^\dag_{21} = (s^\dag_{12})^2 (s^\dag_{34})^2 \nn
s^\dag_{12}s^\dag_{23} s^\dag_{34} s^\dag_{41} + s^\dag_{13}s^\dag_{32} s^\dag_{24} s^\dag_{41} = (s^\dag_{14})^2 (s^\dag_{23})^2 .  \label{eq:S6}
\ea
There are three independent ways of producing a tetramer, represented in the SB language as $s^\dag_{12}s^\dag_{23} s^\dag_{34} s^\dag_{41}|v\rangle$,  $(s^\dag_{12})^2 (s^\dag_{34})^2 |v \rangle $,  and $(s^\dag_{14})^2 (s^\dag_{23})^2 |v\rangle$. All other ways of producing a tetramer, for example, 
$s^\dag_{13}s^\dag_{34} s^\dag_{42} s^\dag_{21} |v \rangle$, become their linear combinations according to the identity (\ref{eq:S6}). 

\begin{figure}[t]
\centering
\includegraphics[width=120mm]{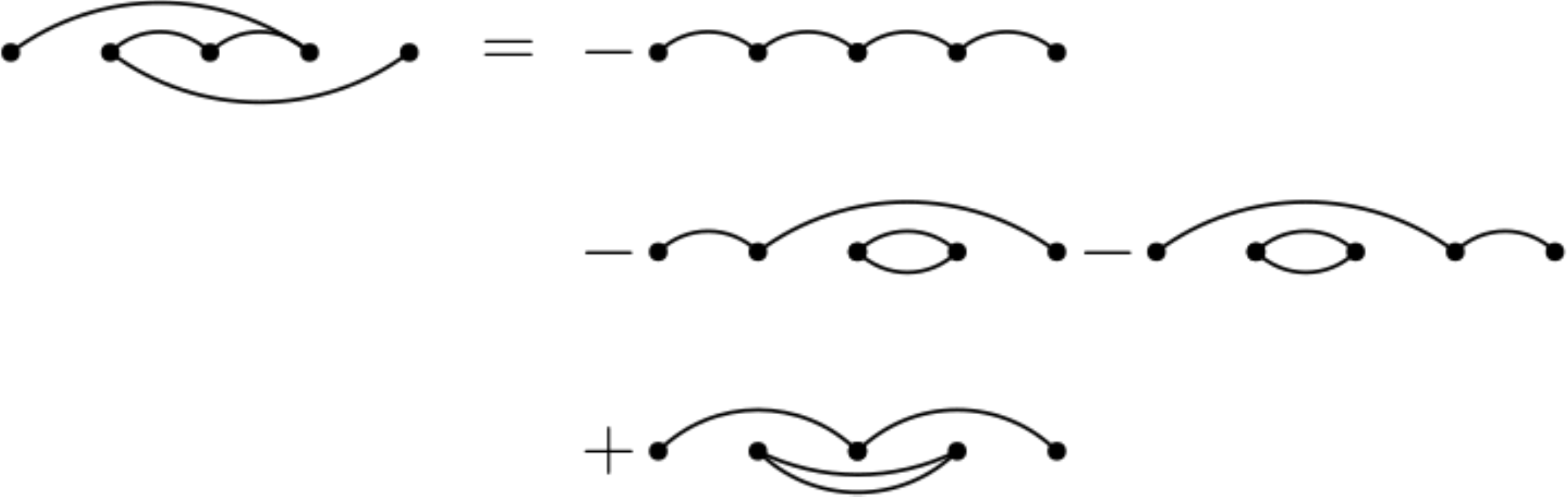}
\caption{Graphical proof showing how the state $|D'\rangle$ breaks down as AKLT state and several one-dimer states.} 
\label{fig:s1}
\end{figure}

One can exploit (\ref{eq:S6}) to prove relations shown graphically in Fig. \ref{fig:s1}. The $|D' \rangle$ state, as mentioned in the main text and Eq. (3), is  generated by the second-order action of the exchange Hamiltonian on the AKLT state. According to our graphical proof, the state $|D'\rangle$ in turn breaks down as a linear combination of the AKLT state, some compact one-dimer states, and the long-ranged one-dimer state. The additional feature of the second-order perturbation is therefore the appearance of the long-ranged dimer, which is realized by the $T_4$ tensor in the main text. 

\section{Spin rotational symmetry}

Our fluctuating $q$-mer MPS ansatz preserves the spin rotational symmetry by remaining in a spin singlet state. In other words, the state transforms trivially under arbitrary global spin rotation $\mathcal{R}(\theta) = \bigotimes_i R_i(\theta) $ with $ R_i(\theta) = e^{i\theta \hat{n}\cdot \vec{S}_i}$ and an arbitrary unit vector $\hat{n}$, i.e., $\mathcal{R} |\psi\rangle = e^{i\phi} |\psi\rangle$. The AKLT state is known to be the spin singlet, and thus the spin singlet nature of our ansatz comes from the local $q$-mers, whose own singlet nature requires proof. In what follows, we show how the bond tensor guarantees the spin singlet nature of the $q$-mers. To this end, we first notice the following relations:
\begin{align}
	R(\theta)_{ik} R(\theta)_{jk'}\left[ {\rm CG}_{1,1}^0\right]_{kk'} = \left[ {\rm CG}_{1,1}^0\right]_{ij}	,\quad
	R(\theta)_{ss'} \left[ {\rm CG}_{1,1}^1\right]_{ij}^{s'} = R^T(\theta)_{ik} R^T(\theta)_{jk'}\left[ {\rm CG}_{1,1}^1\right]_{kk'}^s,
\end{align}
or identically,
\begin{align}
	\includegraphics[width=0.5\textwidth]{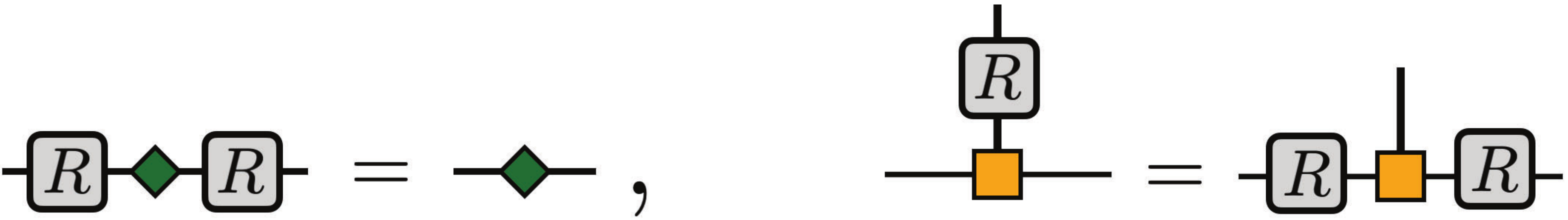}.
	\label{eq:tensor_rotation}
\end{align}
In the first identity, one can consider the CG-matrix as a vector by combining two indices, $\left[ {\rm CG}_{1,1}^0\right]_{ij}  \rightarrow \left[ {\rm CG}_{1,1}^0\right]_{(ij)} $ and then this can be regarded as a wavefunction which is nothing but the singlet wavefunction of two $S=1$ spins. Consequently, it is transformed trivially under the $R(\theta)$-rotation with arbitrary $\theta$. On the other hand, the second equality says that the spin rotation of the resulting spin after the fusion of two spins is identical to the rotation of two spins before the fusion. Using these two identities, one can prove that the $q$-mer with arbitrary length is guaranteed to be the spin singlet. As an example, the tetramer under the spin rotation is shown below:
\begin{align}
	\includegraphics[width=0.4\textwidth]{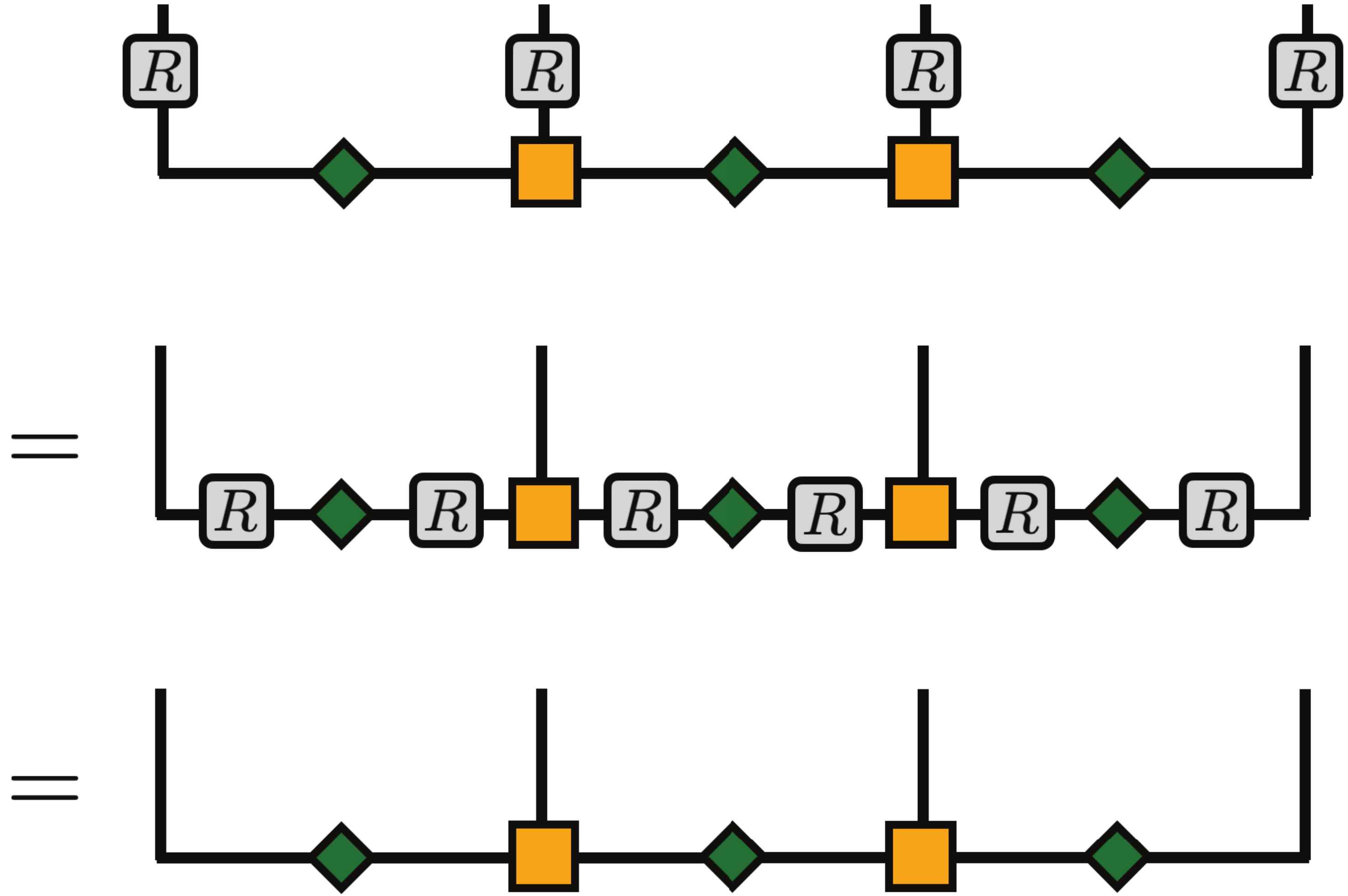}.
\end{align}
Here, the second relation in Eq.\,\eqref{eq:tensor_rotation} is used for the first equality, and the first relation for the second equality above. It clearly shows that $\mathcal{R}(\theta) |{\rm tetramer}\rangle = |{\rm tetramer}\rangle$ regardless of $\theta$ and can be easily generalized to arbitrary $q$-mer.

\section{Identification of Q-mer}

Forming a singlet made out of two or three $S=1$ spins can be done in a unique way, i.e., the definitions of dimer and trimer are unambiguous. On the other hand, there are multiple ways to form the $q$-mers longer than the trimer. In this section, we provide the definition of $q$-mer we generate in this article. First, we note that the CG tensor $[{\rm CG}_{1,1}^1]_{ij}^s$ fuses two $S=1$ spins into $S=1$ spin. In order to visualize this, we assign {\it direction} on each bond of the CG tensor as follows:

\begin{align}
	\includegraphics[width=0.4\textwidth]{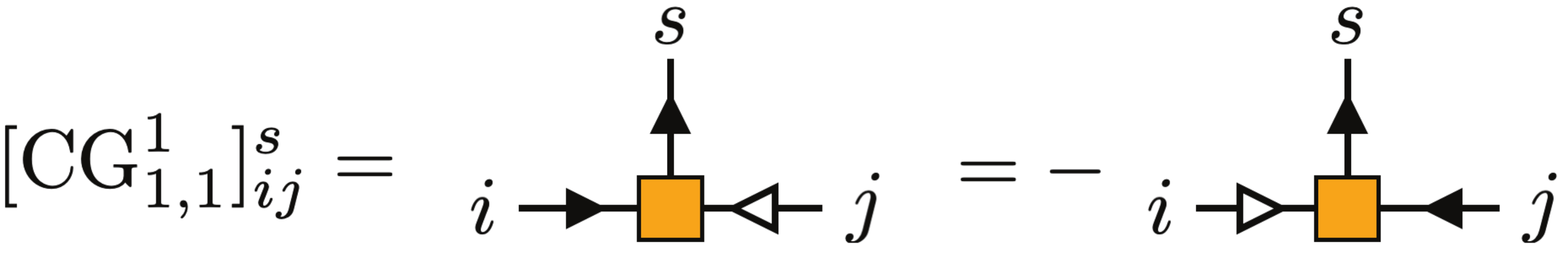},
\end{align}
where two inward directions denote the spins to be fused while the outward one stands for the fused spin. In fact, those directions are related with the quantum number flow\cite{Schollwock11}, particularly $S^z$ quantum number in our ansatz. That is, the sum of the incoming quantum numbers is identical to the outgoing quantum number. The {\it empty} arrow is just to reflect the antisymmetric property of the ${\rm CG}_{1,1}^1$-tensor, i.e., $[{\rm CG}_{1,1}^1]_{ji}^s = -[{\rm CG}_{1,1}^1]_{ij}^s$. We should also notice that the multiplication of the ${\rm CG}_{1,1}^0$ to one of the inward indices in the ${\rm CG}_{1,1}^1$-tensor rotates the quantum number flow, e.g.,

\begin{align}
	[{\rm CG}_{1,1}^0]_{ik} [{\rm CG}_{1,1}^1]_{kj}^s = \frac{1}{\sqrt{3}} \times [{\rm CG}_{1,1}^1]_{si}^j,
\end{align}
or identically

\begin{align}
	\includegraphics[width=0.3\textwidth]{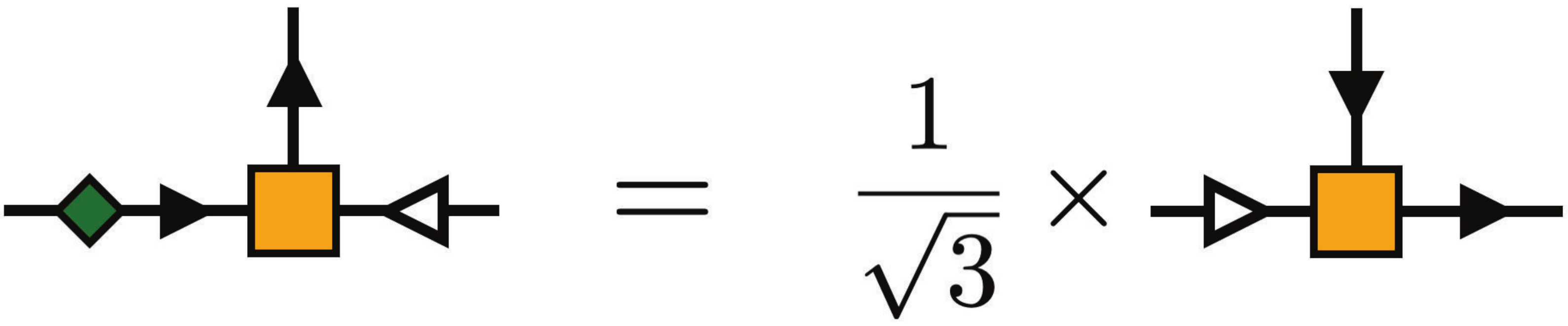}.
\end{align}
Therefore, after multiplying the ${\rm CG}_{1,1}^0$-tensor, the resulting CG-tensor fuses a virtual spin and the physical spin into another virtual spin. Now, as an example, let us consider the 5-mer tensor network: 

\begin{align}
	\includegraphics[width=0.7\textwidth]{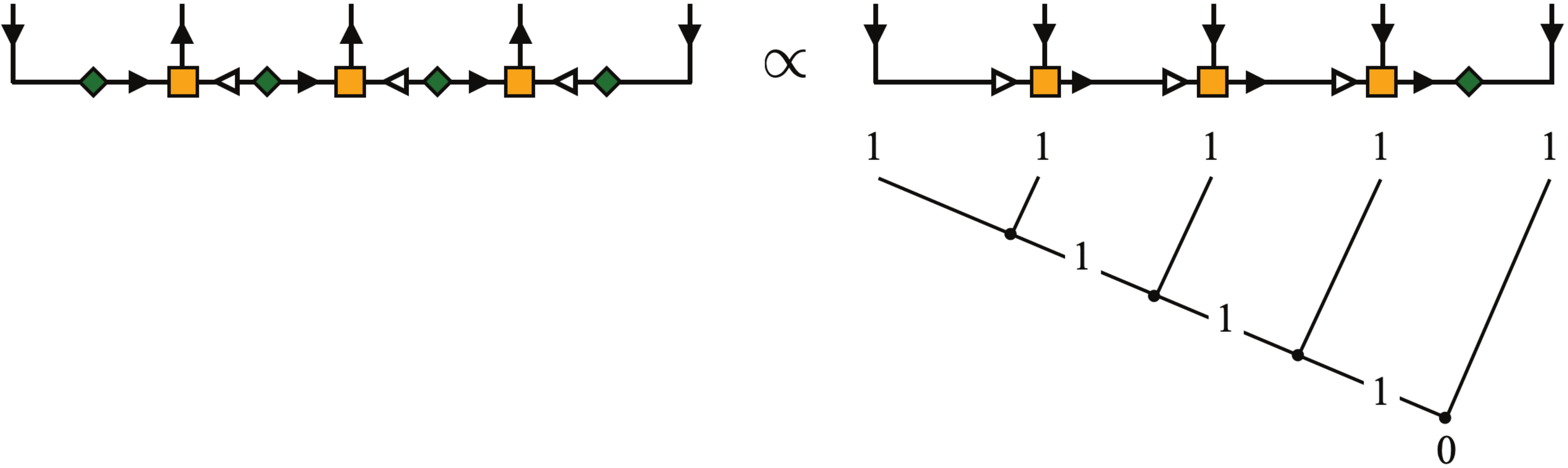}.
\end{align}
By absorbing the bond tensor\,(${\rm CG}_{1,1}^0$ or green diamond) into the site tensor\,(${\rm CG}_{1,1}^1$ or yellow square), the quantum number flows are changed such that two most left (physical) spins are fused into a virtual spin-1 which is fused with third physical spin into another virtual spin-1. In such a way, the physical spins are fused with virtual spin-1's into another virtual spin-1 successively before the last physical spin which combines with the last virtual spin-1 into the spin singlet as depicted in more familiar representation of the spin summation below the right hand side of the above equation. This is more precise definition of $q$-mer in this article.

\section{Further Numerical Results}
\begin{figure}[ht]
	\includegraphics[width=0.75\textwidth]{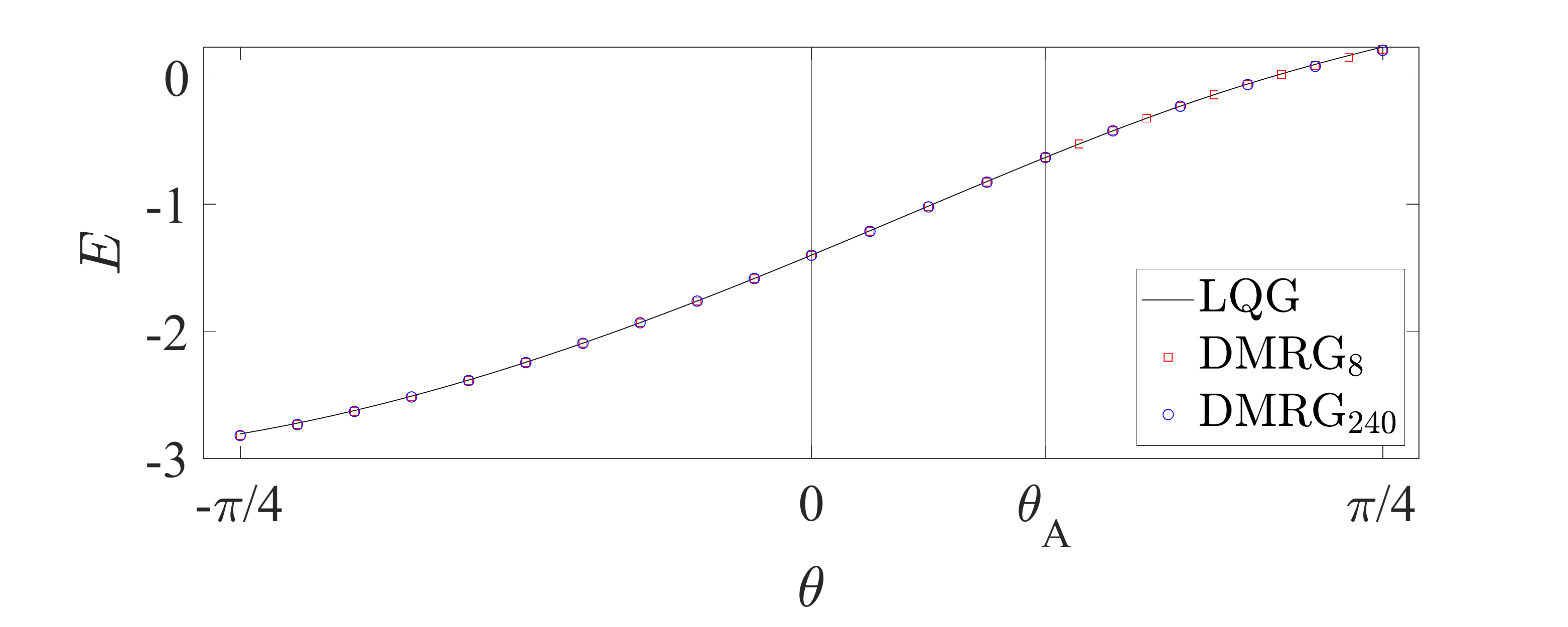}
	\caption{The energy per site obtained by optimized MPS and DMRG.}
	\label{fig:energy}
\end{figure}

The Table I below shows the energy of the optimized MPS state (LQG state) at various $\theta$ values of the BLBQ Hamiltonian over $-\pi/4 \le \theta \le \pi/4$. The DMRG values for the energy are shown for comparison. Two different kinds of runs were done for DMRG, the one where the maximum bond dimension $D$ was fixed to 8 (same as our variational MPS), and where a much bigger maximum bond dimension was used. The energy difference between variational MPS and the DMRG typically occurs in the third significant digit, except when $\theta$ is close to either critical points $\theta = \pm \pi/4$. The DMRG energy was evaluated by taking the average of the local Hamiltonian $H_i = \cos \theta (\v S_i \cdot \v S_{i+1}) + \sin \theta (\v S_i \cdot \v S_{i+1})^2$ for $i$ at the middle of the open chain. The chain size employed in the DMRG was $3\times 10^3$. No size dependence in the energy is discernible at this lattice size. The same energy data is shown as a graph in Fig. \ref{fig:energy}. 

\begin{table}[ht]
\setlength{\tabcolsep}{4pt}
\begin{tabular}{|c|r|r|r|}
\hline
$\theta/\pi $ & \multicolumn{1}{c|}{LQG} & \multicolumn{1}{c|}{$\text{DMRG}_8$} & \multicolumn{1}{c|}{$\text{DMRG}_{240}$} \\ \hline
- 1/4  & -2.8065 & -2.8196 & -2.8286 \\ \hline
-  1/5  & -2.6233 & -2.6297 & -2.6385 \\ \hline
-  3/20 & -2.3831 & -2.3870 & -2.3926 \\ \hline
-  1/10 & -2.0930 & -2.0938 & -2.0981 \\ \hline
-1/20   & -1.7619 & -1.7623 & -1.7640 \\ \hline
\end{tabular}
\quad
\begin{tabular}{|c|r|r|r|}
\hline
$\theta / \theta_A$ & \multicolumn{1}{c|}{LQG} & \multicolumn{1}{c|}{$\text{DMRG}_8$} & \multicolumn{1}{c|}{$\text{DMRG}_{100}$} \\ \hline
0   & -1.4010 & -1.4011 & -1.4015 \\ \hline
1/4 & -1.2090 & -1.2134 & -1.2135 \\ \hline
2/4 & -1.0152 & -1.0213 & -1.0213 \\ \hline
3/4 & -0.8220 & -0.8269 & -0.8269 \\ \hline
1   & -0.6325 & -0.6325 & -0.6325 \\ \hline
\end{tabular}
\quad
\begin{tabular}{|c|r|r|r|}
\hline
$\theta / \pi$ & \multicolumn{1}{c|}{LQG} & \multicolumn{1}{c|}{$\text{DMRG}_8$} & \multicolumn{1}{c|}{$\text{DMRG}_{240}$} \\ \hline
0.1319   & -0.4231 & -0.4231 & -0.4232 \\ \hline
0.1615  & -0.2289 & -0.2293 & -0.2300 \\ \hline
0.1910   & -0.0545 & -0.0559 & -0.0598 \\ \hline
0.2205   & 0.0994 & 0.0893 & 0.0839 \\ \hline
1/4   & 0.2360 & 0.2178 & 0.2098 \\ \hline
\end{tabular}
\label{table:1}
\caption{Ground state energy of the BLBQ Hamiltonian at various $\theta$, where $\theta_A = \tan^{-1}(1/3)$ gives the AKLT point. DMRG$_8$ was done by artificially choosing the maximum bond dimension $D$ to be 8. A much bigger maximum bond dimension $D = 100$ and $D=240$ were employed for DMRG$_{100}$ and DMRG$_{240}$, respectively.}
\label{tab:energy_cmp_lqg_dmrg}
\end{table}

\begin{figure}[ht]
	\includegraphics[width=0.75\textwidth]{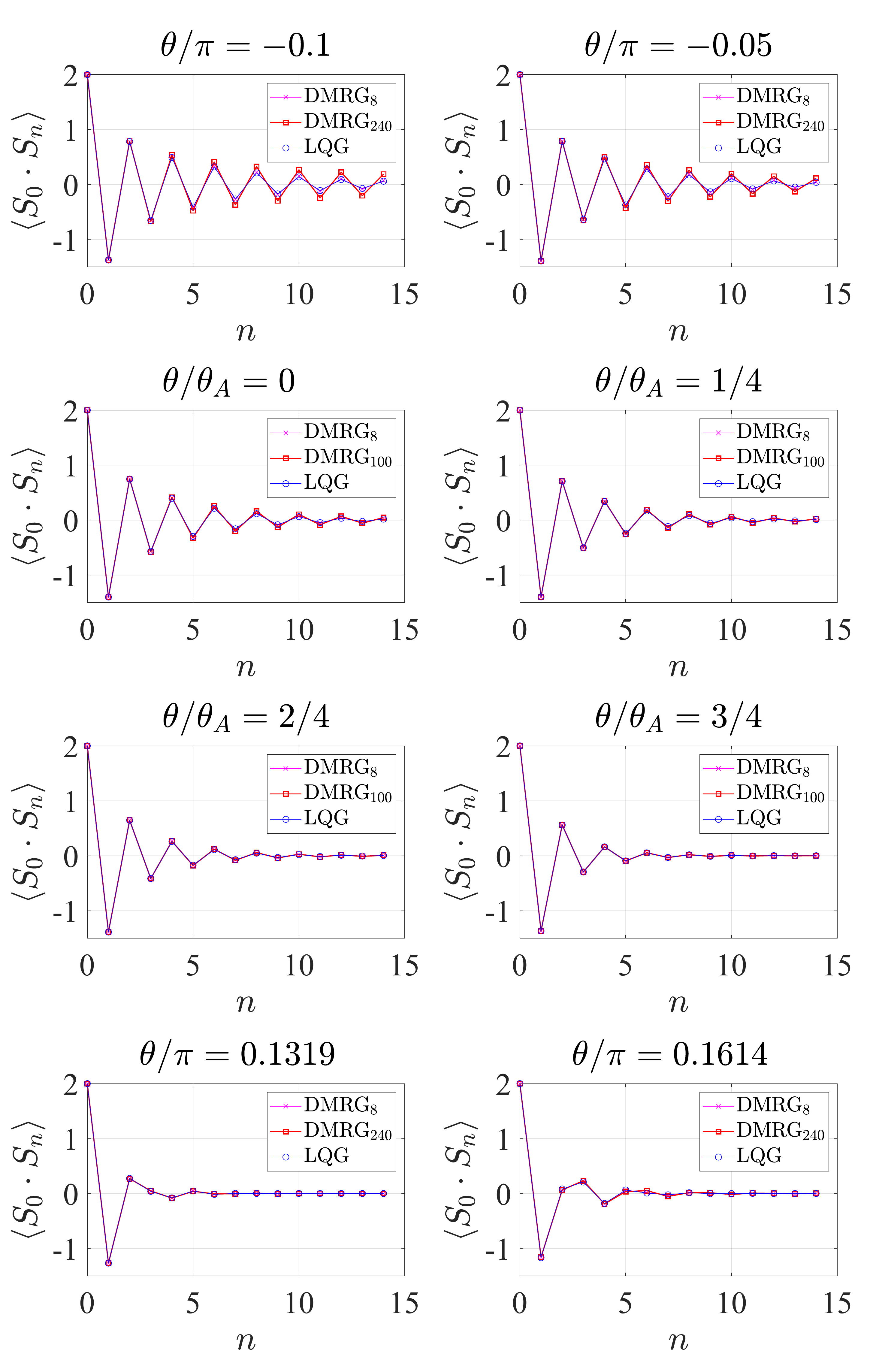}
	\caption{The spin-spin correlation functions obtained by optimized MPS and DMRG.}
	\label{fig:spinspin_vs}
\end{figure}

Figure \ref{fig:spinspin_vs} shows the spin-spin correlation functions obtained by optimized MPS (LQG) and by DMRG at 
several $\theta$ values of the BLBQ model.  


\bibliography{SC}